\documentclass[useAMS,usenatbib]{mn2e}
\usepackage{graphicx}

\title[Effect of Electron Screening on CCSNe]{Effect of Electron Screening  on the Collapsing Process of Core-Collapse Supernovae}
\author[Men-Quan Liu, Ye-Fei Yuan, and Jie Zhang]{Men-Quan Liu$^{1,2}$\thanks{E-mail:
menquan@mail.ustc.edu.cn, liumq@cwnu.edu.cn (MQL);
yfyuan@ustc.edu.cn
(YFY); zhang\_mail@tom.com (JZ)}, Ye-Fei Yuan$^{1}$\footnotemark[1] and Jie Zhang$^{2,3}$\footnotemark[1]\\
$^{1}$Key Laboratory for Research in Galaxies and Cosmology,
University of Science and Technology of China, Chinese Academy of Sciences,\\ Hefei, 230026, China\\
$^{2}$Institute of Theoretical physics, China West Normal
University, Nanchong, 637002, China\\
$^{3}$Institute of Structure and Function, Chongqing University,
Chongqing, 400044, China}
\begin{document}

\date{Accepted  . Received  ; in original form 2008 October 21}

\pagerange{\pageref{firstpage}--\pageref{lastpage}} \pubyear{2008}

\maketitle

\label{firstpage}

\begin{abstract}
By using an average heavy nuclei model, the effects of the electron
screening on electron capture (EC) in core-collapse supernovae are
investigated. A one-dimension code based on the Ws15$M_{\odot}$
progenitor model is utilized to test the effects of electron
screening during the collapsing process. The results show that, at
high densities, the effects of EC on electron
capture becomes significant. During the collapsing stage, the EC
rate is decreased, the collapse timescale is prolonged and the
leakage of the neutrino energy is increased. These effects lead to
an appreciable decrease in the initial energy of the bounce shock
wave. The effects of electron screeening in the other progenitor 
models are also discussed.
\end{abstract}

\begin{keywords}
nuclear reactions, nucleosynthesis, abundances - stars:evolution - supernovae:general.
\end{keywords}

\section{Introduction}

  Supernovae explosion is one of the most violent events in our universe. Their explosion mechanism  is an old
problem, but has not been understood completely ( see e.g. Woosley \& Heger 2007; 
Mart{\'{\i}}nez-Pinedo 2008 for reference). Generally, a
massive star($M\geq10M_{\odot}$) proceeds through all burning stage
from Hydrogen to Silicon, finally leading to an iron core in its
center(Blanc \& Greggio 2008). Electron capture (EC) causes the number
density and the degenerate pressure of electron to decrease with a
lot of neutrino energy loss, which lead to the accelerating collapse
of the star till the central density reaches the maximum (two-three times
the nuclear density). Later on, the infalling outer core collides
with the stiff inner core, and then the bounce shock is produced in
the vicinity of the boundary between the inner core and outer core.
The initial temperature of shock wave is even higher than $10^{11}
\rm K$. At such a high temperature, photon energy  is much larger
than the binding energy of nucleus , and the iron nuclei are
photodisintegrated into protons, neutrons and electrons behind the
shock wave(Janka 2007). Unfortunately with the current numerical
simulations, a self-consistent treatment of one-dimension supernovae
model does not yet lead to successful explosion due to the energy
insufficiency, while two-dimension models show some promise 
(Hix et al. 2003; Burrows et al. 2006).

 It is well known that the weak
interactions, especially the EC and beta decay,
 are essential to the evolution of supernovae. Many basic investigations at
  this aspect have been done by Bahcall(1962,1964), Fuller, Flower, Newmann (1982a;1982b), Aufderheide et al. (1994), Langanke,
   Mart\'{i}nez-Pinedo (1999; 2000) and so on in the last decades.
  Fuller et al. (1982b) have ever mentioned the screening correction but they did not make a detailed calculation. Later,
  Hix \& Thielemann (1996) and Bravo \& Garc\'{i}a-Senz (1999) considered the screening correction on the silicon burning and nuclear
  statistical equilibrium (NSE), and they found that screening effect is
  significant.
  Luo \& Peng (1996) investigated the effect of screening on EC in the supernovae environment by using
  Fuller et al. (1982a,b) method (i.e. so called shell model brink hypothesis). Their results show screening can reduce
  the EC rate by $10-20$ per cent at high density. More recently improved EC rate
  and screening potential were adopted in Liu, Zhang \& Luo (2007).
 The screening
  affects the EC mainly in three aspects. (1) The screening changes the Coulomb
   in-wave function of the electron; however, it can be neglected because the
   screening potential is much less than the average energy of the electrons.
   (2) The screening reduces the energy of the electrons in the capture reaction.
    (3) The screening decreases the number of the high-energy electrons with
     energy greater than the threshold energy of EC. As a result, the reaction threshold
     energy increases. Compared with screening on EC,
  there are at least two differences for beta decay. (1) The electron energy of beta decay is decided by
  the energy difference between parent nucleus and daughter nucleus (including rest mass),
  but the electron energy  in EC can  be much larger than that of the capture threshold. (2)
  In the inner core, beta decay is prohibited due to the Pauli exclusion principle (inhibition degree has been given 
by Peterson \& Bahcall 1963);
  at the outer core, beta decay gets permission, but the  evolution of inner core is quicker than that of outer core at core collapsing stage
  (which is less than 0.5s). So we here neglect the screening effect on the beta decay. Langanke et al. have suggested the screening should be considered in
     the simulations (Langanke \& Mart{\'{\i}}nez-Pinedo 2003), but up to now the screening effect on the total explosion
      process has not been done. Early numerical simulations
        show that the onset of the SN explosion depends on the pre-supernovae model
         and the evolution mode that is especially sensitive to the physical
         parameters  input, which is closely related to the electron fraction and the
   weak interaction rates (Heger, Woosley \& Mart{\'{\i}}nez-Pinedo 2001). Therefore, it is
    imperative to obtain the EC rates with high precision.
As an actual physical input, screening should not be ignored
   since a series of important parameters, such as neutrino energy loss, collapsing timescale etc.,
   is changed. In this paper, we investigate the detailed effects of electron screening on EC in  core-collapse
supernovae.

\section{Pre-supernovae model and computational approach}
We perform numerical simulations by using a modified version of the
one-dimension code developed by Y.R. Wang, S.C. Zhang,  W.Z. Wang, Z.H. Xie
(WZWX  2003) in 1996 . In this code the general
relativistic hydrodynamic equations are adopted (May \& White 1966);
the hydrodynamics method is smoothed particle hydrodynamics (SPH)
referring to Benz (1991); equations of state are similar to Lattimer
\& Swesty (1991) and Cooperstein \& Wambach (1984); the method of neutrino
transport is provided by Suzuki (1994). The pre-supernovae model we
choose is the Ws15$M_{\odot}$ model with an iron core of
1.38$M_{\odot}$ (Woosley \& Weaver 1995). The grid domain includes 1.6
$M_{\odot}$, which is divided into 96 mass layers. Since most of
nuclei involved in supernovae environment are on the unstable and
excited states, it is difficult to get a precise description of each
energy level, especially for heavy nucleus whose excited states are
almost continuous. For a simple consideration, we assumed that the
matter consists of four typical particles including free protons,
free neutrons, $\alpha$-particles, and heavy nuclei, i.e. the so
called `four-particle model'. Such four types of particles can well
represent the whole property of pre-supernovae (Lattimer \& Swesty 1991;
Arcones£¬ Janka \& Scheck 2008). Similar to the other authors, detailed
numerical model indicates the shock is unable to rush out of the
iron core because of too much energy loss in the iron
photodisintegration. So here we mainly investigate the screening
effect in the collapsing process of supernovae explosion.

Usually, the capture rates for the nucleus $(Z, A)$ in thermal
equilibrium at temperature $T$  is given by a sum over the initial
parent states and the final daughter states(Pruet \& Fuller 2003),

\begin{equation}
 \lambda=\sum_{i}\frac{(2J_i+1)e^{-E_i/k_BT}}{G(Z,A,T)}\sum_f\frac{\ln 2}{(ft)_{if}}f_{if}
 \ ,
 \label{1}
\end{equation}
where $J_i$ and $E_i$ are the spin and excitation energy of the
parent states, respectively, $k_B$ is the Boltzmann constant and
$G(Z,A,T)$ the nuclear partition function.  The \emph{ft}-values are
related to {\it GT} and Fermi transition matrix elements. $f_{if}$ is the
phase space integral of electron. But for the `four-particle model'
we adopt, matter is composed of protons, free neutrons,
$\alpha$-particles, and heavy nuclei under 
NSE. Only proton and heavy nuclei can capture
electrons, so the total EC rate
$\lambda=\lambda_{p}+\lambda_{H}$,where $\lambda_{p}$ and $\lambda_{H}$
are the EC rates for proton and heavy nuclei, respectively.
Considering the core of the pre-supernovae is composed of iron
element, the dominant influence of electron screening is decided by
the heavy nuclei. Precise calculation of the EC rate is usually
based on the nuclear shell model, i.e formula (\ref{1}), but the
shortage of energy level quality at high temperature and diverse
electron rates for so many different nucleus will bring complication
and some uncertainty. Therefore, here the Fermi gas model is adopted
to describe the average property of the heavy nuclei. The EC rate is
defined by $\lambda=-(dn_e/dt)$, when the number density of
electron decreases $\lambda>0$; otherwise $\lambda\leq 0$. According
to Bethe et al. (1979), the EC rate for average heavy nuclei is
expressed
\begin{equation}
 \lambda_{\rm H}=1.18\times10^{-44}\frac{3n_{e}}{\mu_{e}^{3}}\frac{3n_{p}}{(p^{p}_{F})^2}\frac{c}{(m_ec^2)^2}m_p~
 \int\!\!\!\int\varepsilon_e^2\varepsilon_\upsilon^2 f_e\
 d\varepsilon_e d\varepsilon_p\ ,
 \label{2}
\end{equation}
where $n_e=\rho N_AY_e,n_p=x\chi_H\rho N_A,
p^p_F=(2m_p\mu_p)^{1/2}$, $\rho$ is the density, $N_A$ is the
Avogardro constant, $\mu_p$ is the chemical potential for proton,
$Y_e$ is the electron fraction, $m_p$ the mass of proton. $x\equiv
Z/A$, $Z,A$ are the mass number and charge of average heavy nuclei,
respectively. $\chi _H$  is the fraction of the average heavy
nuclei, $\varepsilon _e$ and $\varepsilon_p$ is the energy of electron
and proton, respectively.
$f_e=\{1+ \exp[(\varepsilon_e-\mu_e-m_ec^2)/(k_B T)]\}^{-1}$ is
the Fermi-Dirac distribution function for electrons. Because of the
conservation of energy,
$\varepsilon_e+\varepsilon_p=\varepsilon_n+\varepsilon_\upsilon$,
equation (\ref{2}) becomes
\begin{eqnarray}
\lambda_{\rm H} &=& 1.18\times
10^{-44}\frac{3n_e}{\mu_e^3}\frac{3n_{p}}{(p^{p}_{F})^2}m_p
\frac{c}{(m_ec^2)^2} \nonumber \\
&& \times \int_{\mu_p-\Delta_p}^{\mu_p}{\int_{Q}^{\infty}{\varepsilon_e^2(\varepsilon_{e}+\varepsilon_p-\varepsilon_{n})^2f_e
d\varepsilon_{e}d\varepsilon_{p}}}, \label{3}
\end{eqnarray}
where $Q$ is the reaction threshold energy. $Q=|Q_{if}|$ if
$Q_{if}<-0.511$MeV, otherwise $Q=0.511$MeV.
$Q_{if}=(M_pc^2-M_dc^2+E_i-E_f)$, $M_p$ and $M_d$ are the mass of
parent and daughter nucleus, respectively. $E_i$ and $E_f$ the
excited energy for parent and daughter nuclei, respectively. Note
that here parent and daughter nuclei are proton and neutron in bound
state of nucleus, so $Q_{if}= \mu_n +\Delta_n -\varepsilon _p$,
which can also assurance the energy of neutrino is not less than
zero. Proton energy $\varepsilon_p=\mu_p-\Delta_p$, therefore
\begin{eqnarray}
\Delta_p=\mu_p-\varepsilon_p &=& \mu_p-(\varepsilon_n+\varepsilon_{\upsilon}-\varepsilon_e) \nonumber \\
                             &\approx& \mu_p-(\mu_n+\Delta_n+\varepsilon_{\upsilon}-\mu_e)  \nonumber \\
                             &=& \mu_e -\hat{\mu}-\Delta_n-\varepsilon_{\upsilon}=\Delta-\varepsilon_{\upsilon},
\label{l}
\end{eqnarray}
where $\hat{\mu}=\mu_n-\mu_p,
\Delta=\mu_e-\hat{\mu}-\Delta_{n}=1.15\rho_{10}^{-1/8}\sqrt{\mu_e/Y_e}(d\hat{\mu}/d\mu_e)^{1/4}$
is the maximum energy of emissing neutrino, average energy of
neutrino $\bar{\varepsilon_{\upsilon}}\approx (3/5) \Delta$, so
$\Delta_p\approx (2/5) \Delta$.
$\Delta_n=min({3,max(0,\Delta/2))}$, where 3MeV is a refereed value
given by Bethe et al. (1979). Some values at initial moment are listed
in Table 1.

In the high density gas, screening electron cloudy is formed.
Screening reduces the energy of the electron and enhances the
threshold energy, so the EC rate will decrease. We adopt the method
similar to Liu et al. (2007), i.e., we ignore the effect screening
changes the Coulomb in-wave function of the electron, only consider
the effect on the electron energy and threshold energy. So the
capture rate in the strong screening is rewritten as
\begin{eqnarray}
 \lambda_{\rm H}^{'} &=& C\int_{\mu_p-\Delta_p}^{\mu_p}\int_{Q+D}^{\infty}~
 (\varepsilon_e-v_s)^2 \nonumber \\
 &&(\varepsilon_e-\upsilon_s+\varepsilon_p-\mu_n-\Delta_n)^2f_e d\varepsilon_e d\varepsilon_p\,
 \label{3}
\end{eqnarray}
\begin{equation}
C=1.18\times10^{-44}\frac{3n_{e}}{\mu_{e}^{3}}\frac{3n_{p}}{k_{f}^2}\frac{c}{(m_ec^2)^2}
m_p,\
 \end{equation}
 where $D$ is the modification to the
threshold energy, $\upsilon_s$ is the screening potential(Fuller et al. 1982b,
Itoh et al. 2002):
\begin{equation}
 D=2.94\times10^{-5}Z^{2/3}(\rho Y_e)^{1/3} (MeV)\,
 \end{equation}
 \begin{equation}
 v_s=7.525\times10^{-3}Z \left(\frac{Z}{A}\rho_6\right)^{1/3}J (MeV)\,
 \label{vs}
 \end{equation}
 where $\rho_6$ is the density in the unit of $10^6$ g cm$^{-3}$,$J=\sum_{i,j=0}^{10}a_{ij}s^{i}u^{j}$,
 $s=0.5($log$r_s+3),u=1/25(R-25),r_s=1.388\times10^{-2}(A/Z\rho_6)^{1/3}$. $a_{ij}$ can
  be found in Itoh et al. (2002), equation (\ref{vs}) is valid for $10^{-5}\leq r_s\leq10^{-1}$
  (i.e. $\rho \leq 10^{15}$g cm$^{-3}$ ), which is usually fulfilled in the supernovae
  environment.

For the EC on proton, the most probable energy of
interaction $E_0\simeq E_{F}$. Because of $E_0>>E_p$ and $kT>E_p$, where
$E_p$ is the Coulomb energy for proton, strong screening is
invalid and weak screening is valid (Salpeter \& von Horn 1969).
According the method of Kippenhahn \& Weigert (1990),
Bahcall, Chen, \& Kamionkowski (1998) and Kippenhahn 
\& Weigert (1990), weak screening factor is exp$(x_0\pi\eta)$ for $x_0<<1$,
where $x_0=r_c/r_D\simeq (Z_1Z_2/200E_0)(\zeta
\varrho/T_7)^{1/2}$ (note that here $E_0$ is in KeV); $r_c$, $r_D$
are the classical turning-point radius and Debye radius,
respectively; $\zeta=\sqrt{\sum_{i}(X_iZ_i^2/A_i+X_iZ_i/A_i)}$,
$X_i,Z_i$, and $A_i$ are the mass fraction, charge, and mass number,
respectively; $T_7$ is temperature in unit of $10^7$K,
$\eta=(m/2)^{1/2}[(Z_1Z_2e^2)/(\hbar E_0^{1/2})]$, where $m$ is
the reduced mass.  With rough estimate for the core region of
supernova, $T\sim10^9K$, $\zeta\sim10$, and $\rho=10^9\sim
10^{12}\rm g\,cm^{-3}$, weak screening correction is not more than
0.001. On the other hand, proton fraction is much smaller than heavy
nuclei, so screening correction for free proton capture is not
important comparing with that on heavy nuclei.

\section{Simulation Results}
 Because there are many output parameters in the simulation, in Table 1 we
only tabulate some important parameters (only 10 layers are listed)
at the initial moment of simulation (about 0.27s before bounce). One
can see that the mass number and charge of average heavy nuclei,
density, electron fraction, temperature in the pre-supernovae model
and screening potentials at the different layers. Here, $\lambda$
and  $\lambda'$ represent the EC rate without/with screening,
respectively. It is shown in Table 1 that the mass number and charge
of heavy nuclei decrease from the center to the outer layer. The
screening potential is mainly dependent on the density, so both $D$
and $v_s$ decrease monotonically with densities. $\lambda$ is not
only a function of $Z ,A, Y_e$, and $T$, but also a function of
$\lambda _p$,$\lambda_H$,$\chi_p$ and $\chi _H$, so it is not
monotonic to the mass layers. With screening effect into
consideration, it is easy to find that $\lambda'$ is always smaller
than $\lambda$.

\begin{table*}
\centering
 \begin{minipage}{150mm}
 \caption{\label{Tab1}Some important parameters at the initial moment0.27s before bounce.}
\begin{tabular}{l l l l l l l r r r r r r r}
\hline
$j$&$A$&$Z$&$\rho$&$Y_{e}$&$T$&$D$&$v_s$&$\mu_p$&$\mu_n$&$\Delta_p$&$\Delta_n$&$\lambda$&$
\lambda'$\\
$ $ &$ $  &$ $ &$g cm^{-3}$  &$ $ &$K$ &$MeV$ &$MeV$ &$MeV$ &$MeV$
&$MeV$ &$MeV$
 &$s^{-1}cm^{-3}$ &$s^{-1}cm^{-3}$\\\hline 1   & 61.6 & 26.1
&8.3E+09 &   0.42 & 8.1E+09 & 0.39& 0.14    & -12.54 & -6.72
&   0.37    &   0.92    & 6.28E+30 & 5.41E+30\\
 10 &  60.1 &   25.8
& 5.3E+09 & 0.43 & 7.9E+09 & 0.33 &   0.12    & -12.04  &   -7.06 &
0.32    & 0.79    & 1.86E+30    &   1.65E+30\\
 20  &   58.6    &
25.4 & 3.4E+09 & 0.43 &   7.7E+09 &   0.29    & 0.10    & -11.47 &
-7.46
& 0.33    & 0.82    &   1.13E+30    & 1.02E+30\\
 30  & 57.1 & 25.1 &
2.2E+09 &   0.44    &   7.5E+09 & 0.25 &   0.08 & -10.83 & -7.93 &
0.39    &   0.98    & 1.14E+30 & 1.05E+30\\
 40  &   55.0 &   24.6 &
1.3E+09 &   0.45 & 7.3E+09 & 0.21    &   0.07 & -9.95   & -8.60   &
0.54 & 1.35 & 1.71E+30    & 1.59E+30\\
 50  & 52.9    & 24.1 &
7.5E+08 &   0.46 &   6.8E+09 & 0.17    &   0.06 & -9.18   & -9.18 &
0.66 & 1.65    & 1.50E+30    & 1.40E+30\\
 60 & 52.1    & 23.9    &
3.4E+08 &   0.46 & 7.1E+09 &   0.13    & 0.04    &   -9.19 & -9.19 &
0.47 & 1.17    &   1.80E+29 & 1.73E+29\\
 70  &   49.6 & 23.2 &
1.2E+08 & 0.47    & 6.1E+09 & 0.09    &   0.03    & -9.13   & -9.13
& 0.31 & 0.77 & 7.20E+27 & 6.90E+27\\
 80 &   43.8    & 21.8 &
2.7E+07 & 0.50 & 4.7E+09 & 0.06    & 0.02    &   -8.89 & -8.89 &
0.17 & 0.41 & 9.08E+25 & 8.77E+25\\
 90  & 43.3    & 21.3 & 2.0E+06 &
0.50    & 2.9E+09 & 0.02    & 0.01    & -8.80 & -8.80 &   0.04    &
0.11 & 3.57E+22 & 3.48E+22\\

 \hline
\\\end{tabular}
{\it note.} j denotes the  layer number.
 \end{minipage}
\end{table*}

\begin{figure}
 \includegraphics[width=0.4\textwidth]{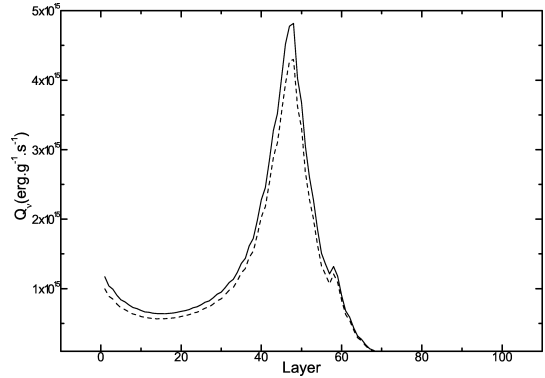}
  \caption{Neutrino energy loss rates  at 0.27s before bounce. Solid curve is
neutrino energy loss rate without screening, while dashed curve is
that with screening in the different layers.} \label{fig1}
\end{figure}

\begin{figure}
 \includegraphics[width=0.4\textwidth]{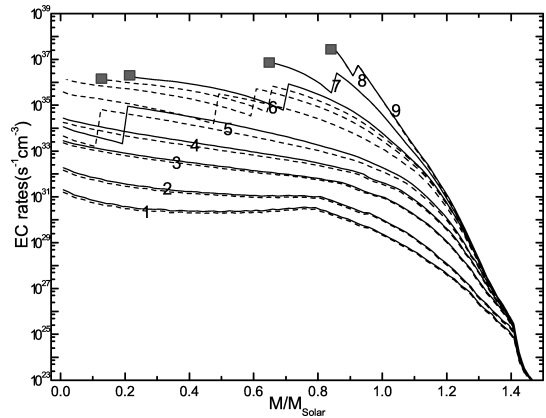}
\caption{The electron capture rate at different collapsing time.
Solid curves are EC rates without screening, while dashed curves are
that with screening in the different layers.The symbols 1, 2, 3, 4,
5, 6, 7,8, 9 represent the time of -250.0, -200.0, -100.0,
-50.0, -25.0, -5.0, -1.5, 0 and 6ms for the case of
without screening ( $'-'$ means the time is before bounce, ms is the
abbreviation for milliseconds). Squares denote the neutrino trapping
positions at the corresponding time.}
 \label{fig2}
\end{figure}

\begin{figure}
 \includegraphics[width=0.4\textwidth]{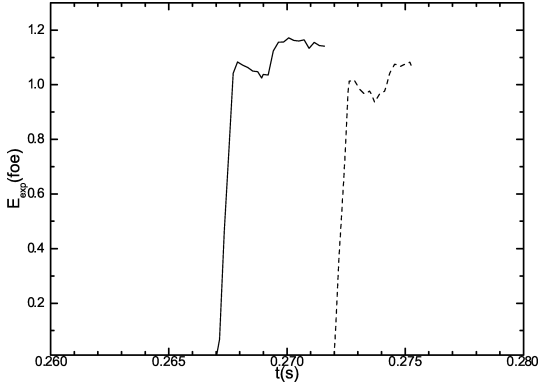}
 \caption{The explosion energy with/without screening versus collapsing time. The notes are the same as those in Fig.1.
 }
 \label{fig3}
\end{figure}

\begin{figure}
 \includegraphics[width=0.4\textwidth]{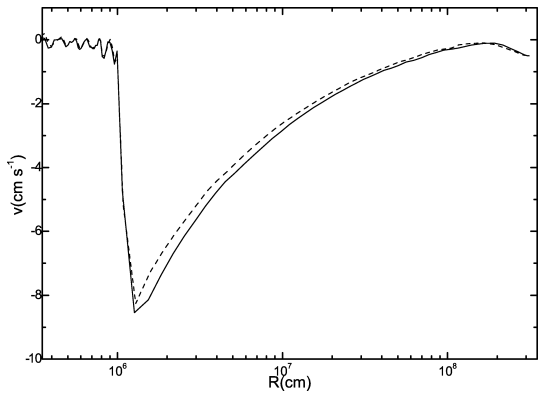}
 \caption{The collapsing velocity versus radius at bounce. Shock is  produced  at sound speed position, and radius of PNS  appears at the vicinity of maximal collapsing velocity at the bounce.The notes are the same as those in Fig.1. }
 \label{fig3}
\end{figure}

The neutrino leakage and diffusion process are the key factors on
the explosion energy, and the energy loss of neutrino exceeds
 90 per cent of the total in the whole evolution process of the star.
 Neutrino energy loss rate is an important and complicated parameter
 to affect the explosion energy and evolution, which is related with
 the weak interaction rate, neutrino energy, and transport equation
 and so on. The change of EC rate must influence the neutrino energy
 loss rate. Fig. \ref{fig1} shows the comparison of neutrino loss rate
 with/without screening at 0.27s before bounce. Solid curve denotes
 loss rate without screening, dotted curve denotes that with screening.
  We find neutrino energy loss rate decreases generally, at some
  region the modification is more than $5$ per cent (It closely depends on EC
 rate),   but for outer part of iron core, it has hardly changed.

Fig. \ref{fig2} shows the evolution of EC rate with/without
screening at different moment during the  collapsing  stage.
   One can find (i) the EC rate with screening is always smaller than that without screening, 
   which is caused by the decrease of the EC rate and the delay of collapse
   time-scale. At the initial stage (such as symbol 1), the difference is
   mainly caused by the different EC rates, and at the later stage (such as symbols
   5,6,7), the difference comes from a different collapse history, while at
   bounce moment, for the case without screening(symbol 8), the EC rates are similar to that at moment 6ms later for the case with screening.
   (ii) In the inner part of the core, both the EC rates and the difference between the case with screening and the case without screening
   increase rapidly with time. The reason is that the densities increase quickly  with the
   collapsing process.  The higher the density, the larger the screening potential. 
   However the screening effect is decided not only
   by the
   potential, but also by the Fermi energy of electrons. When the density is
   large enough, the Fermi energy becomes much larger than the potential energy.
  Correspondingly the screening effect will decrease.(iii) In the outer part of the core, the EC rates change comparatively slow.
  At the edge of iron core ($1.38 M_{\odot}$), there is an inflexion because both the fraction
  of heavy nuclei and the densities break obviously. At the outer envelop, EC rate is almost stable since the time is just 0.3s.
  (iv) Note the squares symbols in Fig. 2, from left to right, they denote  neutrino trapping critical position
  at  -5.0, -1.5 and 0.0ms before bounce, respectively. We can find the high
  density region extents rapidly with time. Because of the inverse process of
  EC rate $[\upsilon_e+(A,Z)\rightleftharpoons e^{-}+(A,Z+1)]$ enhances with  density, above critical density
($\rho_{trapping}$   is about $3\times  10^{12} g cm^{-3}$),
  an equilibrium of neutrino will be  reached and the lepton fraction (including
    electron fraction and neutrino fraction) will keep as a constant.
Surely, at such situation our method for screening modification is
invalid to change of lepton fraction, but screening is still
exist and keep effect in the reversible
    reaction.

    We also compare the collapsing time and the radius of protoneutron star (PNS) with
    screening/without screening(as shown in Fig. 3 and 4). We find (i) the collapsing
    time is prolonged; t=0.267s without screening while t=0.272s
    with screening. The reason is that screening decreases the capture rate.
    This makes the electron number density and degenerate pressure to drop
    a little more slowly than the fiducial case. During the collapsing stage,
    the total pressure is dominated by the electron degenerate pressure,
     so that the decrease of total pressure also become more slowly and
     the collapsing velocity decreases comparatively. (ii) The initial energy of shock wave decreases. The initial energy
     of shock wave at about 0.8ms after bounce is 1.06 foe (1foe=$1\times 10^{51}erg$)
     without screening and 1.01foe with screening. Their difference is 0.05 foe, which is 5 per cent of the total.
    On the one hand, it could be that the calculation with screening results in a different radial
      profile of the star at bounce when compared with a calculation without screening.
      The different profile could result in differences in the propagation of the shock wave.
     Figure 4 shows that the velocity of the collapsing matter is different for radii above the shock
     position;
       on the other hand, it could be the total neutrino energy loss increase due to the time delay. The total neutrino energy loss
     $\dot{Q}_{\upsilon t}=\sum _{i}\sum_{j}\dot{Q}_{\upsilon}\Delta M_{j}\Delta t$ ,
 where $t$ is the time, $\Delta t$ is the time step, $\Delta M_{j}$ is the mass
 of  the $jth$ layer. Despite the screening effect decreases the capture
 rate and the neutrino energy loss rate, the total energy loss is also related
 to the collapsing time.(iii) Radius of PNS
     is hardly changed, but the radius of outer part of iron core increase
     slightly. It is favorable to the success of supernovae explosion.

  \section{Conclusion}
  In this paper, we have considered screening effect on EC, the corresponding
   effects on the collapsing process, and on the initial shock wave energy of core-collapse supernovae.
   Our results show that the screening effect is appreciable. Since there are both advantages
    and disadvantages to the final explosion, more detailed simulation is needed by using
    more concrete equation of state and progenitor model (Hix et al. 2003).
In order to investigate the exact effect of screening, the concrete
nuclide and the other methods are needed. In recent years, many
methods such as the large-scale shell model and the pn-random phase
approach are widely investigated (Nabi \& Sajjad 2008). Improved results
are universally one order of magnitude smaller than Fuller et al. (1982a,b)
results. If
the improved EC rates are considered, the screening
effect should increase comparatively. Furthermore, for two or three
dimension numerical simulations of core-collapse supernovae, our
method for electron screening is also valid. The explosion mechanism
of the core-collapse supernovae has been investigated extensively in
the last  four decades and significant progress has been obtained,
but some of the most fundamental questions are still unanswered. It
is necessary to include the effect of screening in later detailed
simulations.

\section*{Acknowledgments}
We are grateful to  the referee for her/his valuable suggestions to
improve our manuscript. We also thank Professor S.E. Woosley for
providing us with his progenitor data. M-QL
would like to thank WZWX for providing the original program to
simulate the supernova explosion. This work is partially supported by
National Basic Research Program of China (grant 2009CB824800), the
National Natural Science Foundation (grant
10733010,10673010,10573016), and Program for New Century Excellent
Talents in University.This work is also supported by the Youth Fund
of Sichuan Provincial Education Department (grant 2007ZB090) and the
Science and Technological Foundation of China West Normal
University(grants 07A004, 07A005)

\end{document}